\documentclass[10pt]{article}
\usepackage{graphicx}

\def\Title#1{\begin{center} {\Large #1 } \end{center}}
\def\Author#1{\begin{center}{ \sc #1} \end{center}}
\def\Address#1{\begin{center}{ \it #1} \end{center}}

\newcommand\pubblock{\rightline{\begin{tabular}{l} Proceedings of the Second Annual LHCP\\ \pubnumber\\
         \pubdate  \end{tabular}}}

\newenvironment{Abstract}{\begin{quotation} \begin{center} 
             \large ABSTRACT \end{center}\bigskip 
      \begin{center}\begin{large}}{\end{large}\end{center} \end{quotation}}

\newenvironment{Presented}{\begin{quotation} \begin{center} 
             PRESENTED AT\end{center}\bigskip 
      \begin{center}\begin{large}}{\end{large}\end{center} \end{quotation}}

\def\beq{\begin{equation}}
\def\eeq#1{\label{#1}\end{equation}}
\def\eeqn{\end{equation}}

\def\beqa{\begin{eqnarray}}
\def\eeqa#1{\label{#1}\end{eqnarray}}
\def\eeqan{\end{eqnarray}}

\let\bar=\overbar

\def\Dslash{\not{\hbox{\kern-4pt $D$}}}
\def\dslash{\not{\hbox{\kern-2pt $\del$}}}

\def\msb{{\bar{\ssstyle M \kern -1pt S}}}

\usepackage{booktabs}
\usepackage{color}
\usepackage{amsmath}
\usepackage{lineno}
\usepackage{float}

\newcommand\pubnumber{ATL-PHYS-PROC-2014-107 } 

\newcommand\pubdate{\today}

\def\affiliation{
On behalf of the ATLAS Collaboration, \\
Max-Planck-Institut f\"ur Physik \\
F\"ohringer Ring 6, 80805 Munich, Germany }

\begin{document}
\large
\begin{titlepage}
\pubblock

\vfill
\Title{  Search for Supersymmetry in the
4-lepton final state with the ATLAS
detector
  }
\vfill

\Author{ Maximilian Goblirsch-Kolb }
\Address{\affiliation}
\vfill
\begin{Abstract}
The four-lepton final state, rare in standard model processes, is an attractive channel
for new physics searches at the LHC. In supersymmetry (SUSY), charged leptons may
arise from R-Parity conserving (RPC) cascade decays of SUSY particles to the lightest supersymmetric particle,
or from R-Parity violating (RPV) decays to standard model particles. This paper presents
a search for supersymmetry in events with at least four charged leptons carried out
using 20.3~$\text{fb}^{-1}$ of proton-proton collision data taken with the ATLAS detector at $\sqrt{s} =$~8 TeV during the 2012 LHC run. Up to two of the leptons may be hadronically
decaying tau leptons, enhancing the sensitivity to tau-enriched scenarios. A high
sensitivity to a wide range of R-Parity conserving and violating supersymmetric models
is achieved.

\end{Abstract}
\vfill

\begin{Presented}
The Second Annual Conference\\
 on Large Hadron Collider Physics \\
Columbia University, New York, U.S.A \\ 
June 2-7, 2014
\end{Presented}
\vfill
\end{titlepage}
\def\thefootnote{\fnsymbol{footnote}}
\setcounter{footnote}{0}
%

\normalsize 


\section{Introduction}
This paper summarizes a search presented in detail in~\cite{Aad:2014iza}, which targets signatures with at least four charged leptons. 
The analysis uses the full dataset of proton-proton collisions collected by the ATLAS detector \cite{Aad:2008zzm} during the 2012 LHC run at a center-of-mass energy of $\sqrt{s} $= 8 TeV, corresponding to an integrated luminosity of $\int L dt = 20.3$~$\text{fb}^{-1}$. 

\begin{figure}
    \includegraphics[width = 0.32 \textwidth]{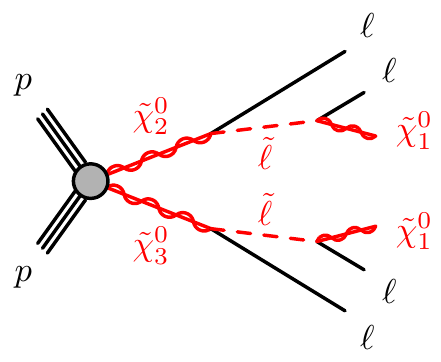}
    \includegraphics[width = 0.32 \textwidth]{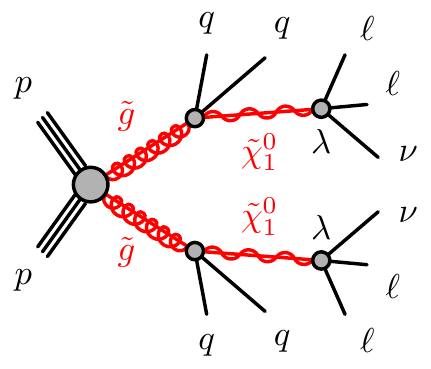}
    \includegraphics[width = 0.32 \textwidth]{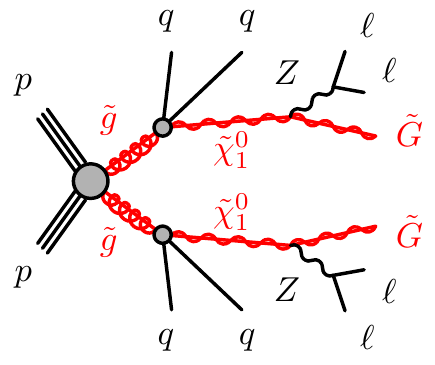}
    \caption{Example of signal processes leading to a final state with four charged leptons: RPC Neutralino pair production (left), RPV neutralino decays after gluino pair production (center), a GGM model with a neutralino NLSP\cite{Aad:2014iza} \label{FGs}}
\end{figure}

\section{Signal Selection}
\begin{table}[H]
\vspace{-0.2in}
\centering
 \caption{The selection requirements for the signal regions, where $\ell=e,\mu$ and \emph{SFOS} indicates two same-flavor opposite-sign light leptons. The invariant mass of the candidate Z boson in the event selection can be constructed using two or more of the light leptons present in the event: all possible lepton combinations are indicated for each signal region \cite{Aad:2014iza}.\label{SRtable}}
\small{
 \begin{tabular}{l c c c c c c }
\hline\hline
            & ~~~N($\ell$)~~~ & ~~~N($\tau$)~~~ & Z-boson veto & $E_{T}^{\text{Miss}}$~[GeV] & & $m_{\text{eff}}$~[GeV]   \\ 
\hline
SR0noZa  & $\geq$4 & $\geq$0 & SFOS, SFOS+$\ell$, SFOS+SFOS & $>$50 & & -- \\
SR1noZa    & $=$3 & $\geq$1  & SFOS, SFOS+$\ell$ & $>$50 & & -- \\
SR2noZa    & $=$2 & $\geq$2  & SFOS & $>$75 & & -- \\
\hline
SR0noZb     & $\geq$4 & $\geq$0 & SFOS, SFOS+$\ell$, SFOS+SFOS & $>$75 & or & $>$600 \\
SR1noZb    & $=$3 & $\geq$1     & SFOS, SFOS+$\ell$ & $>$100 & or & $>$400 \\
SR2noZb    & $=$2 & $\geq$2     & SFOS & $>$100 & or & $>$600 \\
\hline\hline
            & N($\ell$) & N($\tau$) & Z-boson requirement & $E_{T}^{\text{Miss}}$~[GeV] & &   \\ 
\hline
SR0Z     & $\geq$4 & $\geq$0 & SFOS & $>$75   &  & --  \\
SR1Z     & $=$3    & $\geq$1 & SFOS & $>$100  &  & --  \\
SR2Z     & $=$2    & $\geq$2 & SFOS & $>$75   &  & --  \\
\hline\hline
 \end{tabular}
}
\end{table}
The main requirement on events is the presence of at least four isolated charged leptons. 
Leptons are classified as \emph{light leptons} (e,$\mu$) and hadronically decaying $\tau$-leptons. At least two of the four leptons must be light leptons.
To  further select Signal events and reject potential standard model background, the properties of the expected supersymmetric signal are used.
The missing transverse energy $E_{T}^{\textbf{Miss}}$ is expected to be high in the case of R-Parity conserving models, where the lightest supersymmetric particle (LSP) escapes the detector.
In case of a light LSP or R-Parity violation, the effective mass,
$$ m_{\text{eff}} = \sum\limits_{\text{Leptons}} p_{T} + \sum\limits_{\text{Jets}} p_{T} + E_{T}^{\text{Miss}}$$
can instead be used as a powerful means of discrimination. It is sensitive to the mass scale of the initially produced supersymmetric particles.
Finally, the presence or absence of a Z boson decay (2, 3 or 4 light leptons with a combined invariant mass within 10 GeV of the Z resonance) is used to enhance sensitivity to signal models without Z boson emission.
This results in a total of nine signal regions, as presented in Table~\ref{SRtable}.

\section{Standard Model background}

The standard model background is classified into two categories. 
\begin{itemize}
    \item Processes with at least four prompt leptons are called \emph{Irreducible} and estimated using Monte-Carlo simulation.
The main contributions of this kind are from ZZ, $t\bar{t}$Z, Higgs and Triboson events. 
The uncertainty on these backgrounds is dominated by the theoretical prediction of the observables used in the analysis. 
\item Process with fewer than four prompt leptons are called \emph{Reducible}. Here, additional leptons (\emph{fakes}) may come from secondary decays or misidentifications. For example, a QCD jet may fake a hadronic $\tau$ candidate. This background is mainly made up of WZ, $t\bar{t}$ and Z+jets events. As the behavior of these fake leptons is not perfectly predicted by the simulation, a data-driven method is used to estimate these backgrounds. The dominant uncertainty on this background contribution is the attainable accuracy of this method. 
\end{itemize}
In signal regions with four light Leptons, the irreducible background is dominant. As soon as hadronic $\tau$ decays are selected, this changes and the reducible component gains importance. This due to the difficulty in separating hadronic $\tau$ decays from other hadronic jets.

\section{Results}

\begin{table}[htb]
 \centering
  \caption{The number of data events observed in each signal region, together with background predictions in the same regions. 
  Quoted uncertainties include both the statistical and systematic uncertainties, taking into account correlations. 
Where a negative uncertainty reaches down to zero predicted events, it is truncated \cite{Aad:2014iza}.
\label{tab:SRdata}}
\footnotesize
\renewcommand\arraystretch{1.3}
\begin{tabular}{ c   ccccccc | c }\hline
   & ZZ & tWZ & $t\bar{t}$Z & VVV & Higgs & Reducible & $\Sigma $ SM 
& Data 
\\
\hline\hline
SR0noZa & $0.29\pm0.08$  & $0.067\pm0.033$  & $0.8\pm0.4$  & $0.19\pm0.09$  & $0.27\pm0.23$  & $0.006^{+0.164}_{-0.006}$  & $1.6\pm0.5$  
& $3$ 
\\
SR1noZa & $0.52\pm0.07$  & $0.054\pm0.028$  & $0.21\pm0.08$  & $0.14\pm0.07$  & $0.40\pm0.33$  & $3.3^{+1.3}_{-1.1}$  & $4.6^{+1.3}_{-1.2}$  
& $4$ 
\\
SR2noZa & $0.15\pm0.04$  & $0.023\pm0.012$  & $0.13\pm0.10$  & $0.051\pm0.024$  & $0.20\pm0.16$  & $3.4\pm1.2$  & $4.0^{+1.2}_{-1.3}$  
& $7$ 
\\
SR0noZb & $0.19\pm0.05$  & $0.049\pm0.024$  & $0.68\pm0.34$  & $0.18\pm0.07$  & $0.22\pm0.20$  & $0.06^{+0.15}_{-0.06}$  & $1.4\pm0.4$
& $1$ 
\\
SR1noZb & $0.219^{+0.036}_{-0.035}$  & $0.050\pm0.026$  & $0.17\pm0.07$  & $0.09\pm0.04$  & $0.30\pm0.26$  & $2.1^{+1.0}_{-0.9}$  & $2.9^{+1.0}_{-0.9}$
& $1$  
\\
SR2noZb & $0.112^{+0.025}_{-0.024}$  & $0.016\pm0.009$  & $0.27^{+0.28}_{-0.27}$  & $0.040\pm0.018$  & $0.13\pm0.12$  & $2.5^{+0.9}_{-1.0}$  & $3.0\pm1.0$
& $6$  
\\
\hline
SR0Z & $1.09^{+0.26}_{-0.21}$  & $0.25\pm0.13$  & $2.6\pm1.2$  & $1.0\pm0.5$  & $0.60^{+0.22}_{-0.21}$  & $0.00^{+0.09}_{-0.00}$  & $5.6\pm1.4$
& $7$  
\\
SR1Z & $0.59^{+0.11}_{-0.10}$  & $0.042\pm0.022$  & $0.41\pm0.19$  & $0.22\pm0.11$  & $0.14\pm0.05$  & $1.0\pm0.5$  & $2.5\pm0.6$ 
& $3$  
\\
SR2Z & $0.70^{+0.12}_{-0.11}$  & $0.0018\pm0.0015$  & $0.035\pm0.024$  & $0.039\pm0.014$  & $0.14^{+0.04}_{-0.05}$  & $0.9\pm0.5$  & $1.8\pm0.5$
& $1$  
\\
\hline
\end{tabular}
\end{table}

Table \ref{tab:SRdata} shows a breakdown of the expected and observed event counts in the various signal regions.
No significant excess over the standard model prediction is observed. 
This result is interpreted in a set of simplified RPC and RPV models, as well as in the framework of GGM SUSY. 

\begin{figure}[htb]
\vspace{-1.6in}
    \includegraphics[width = 0.48 \textwidth]{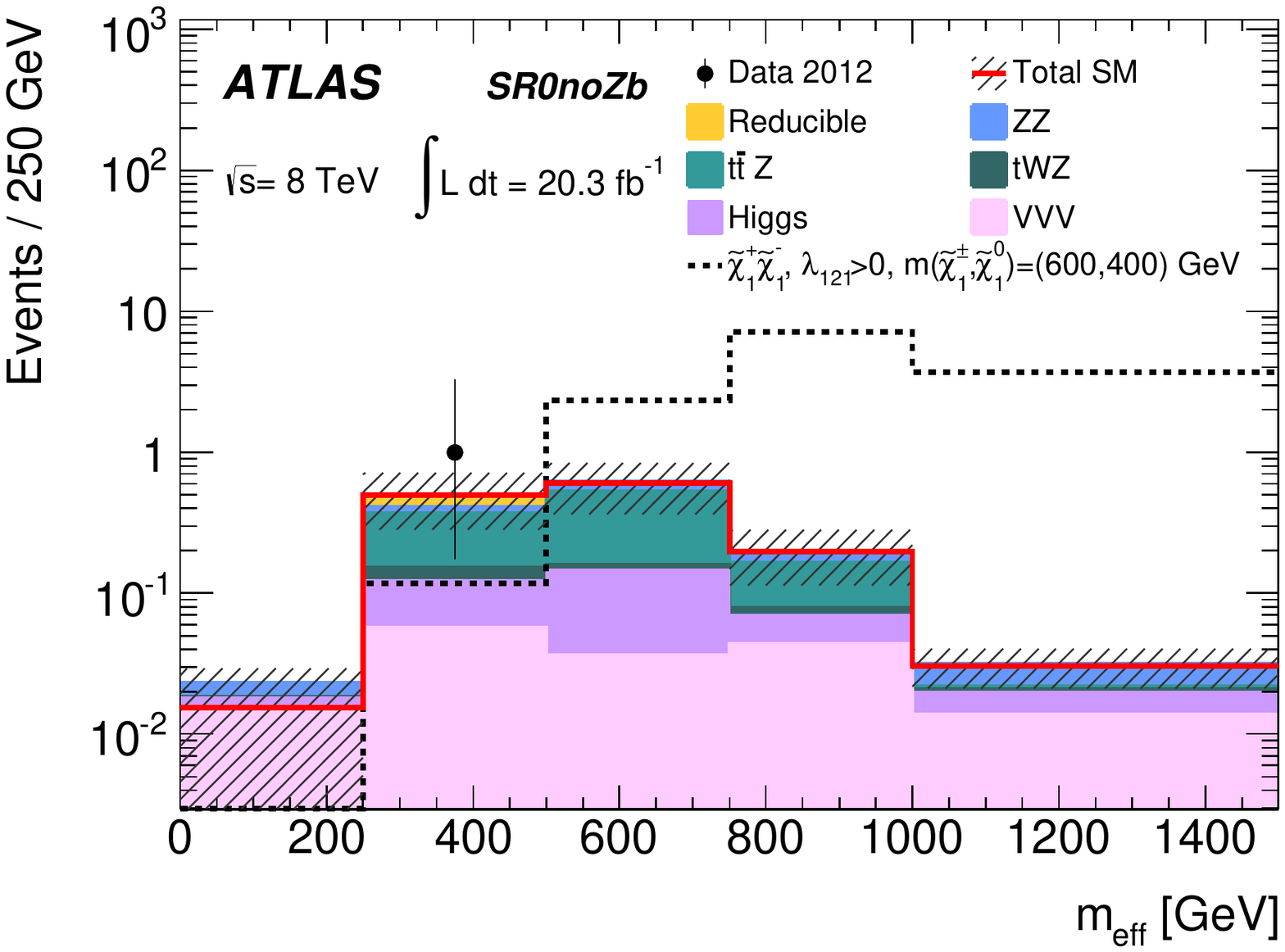}
    \includegraphics[width = 0.48 \textwidth]{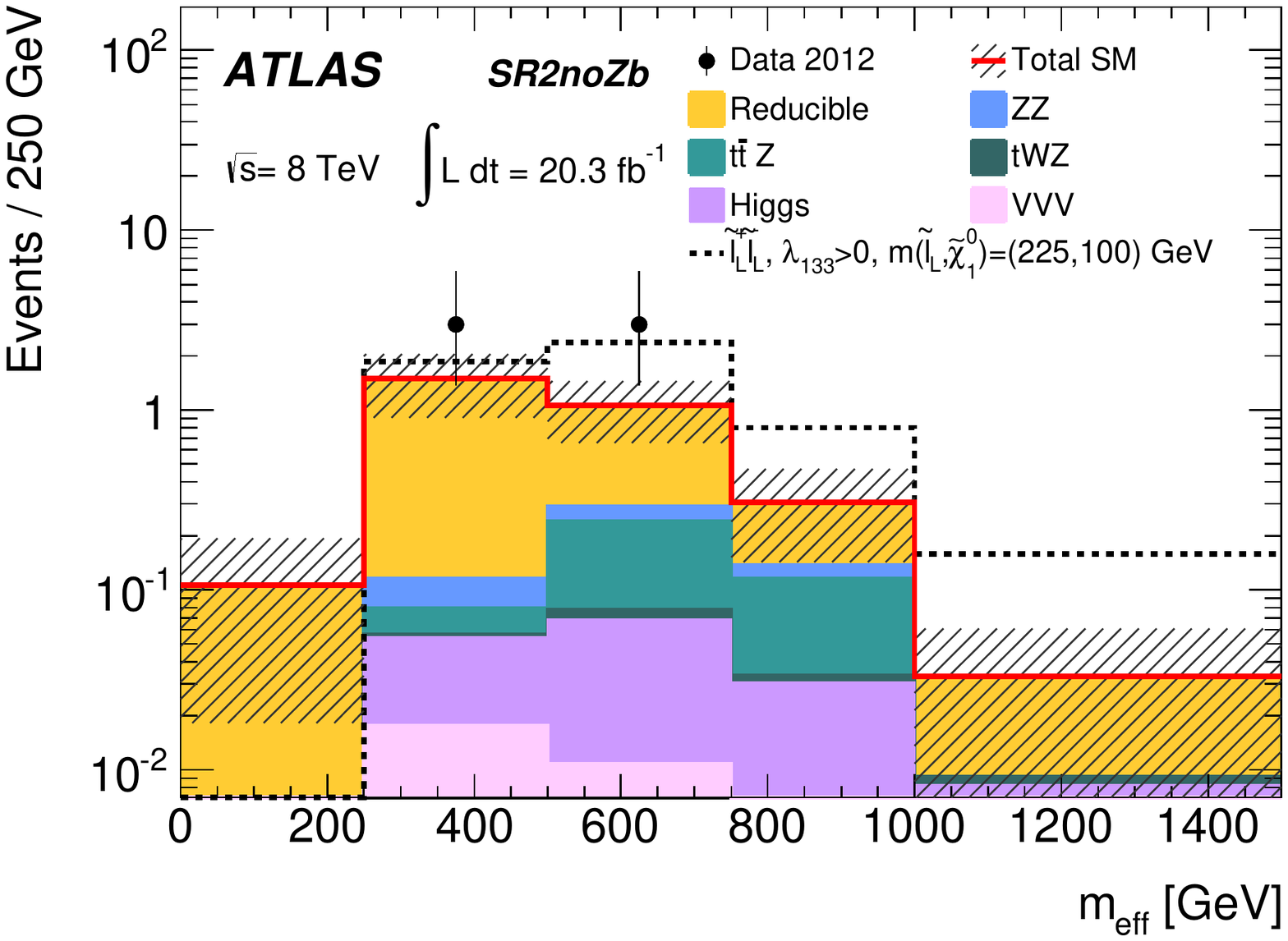}
    \caption{Effective mass distribution in the SR0noZb (left) and SR2noZb (right) signal regions\cite{Aad:2014iza}. \label{SRplots}}
\end{figure}

\begin{figure}[htb]
\vspace{-1.6in}
    \includegraphics[width = 0.39 \textwidth]{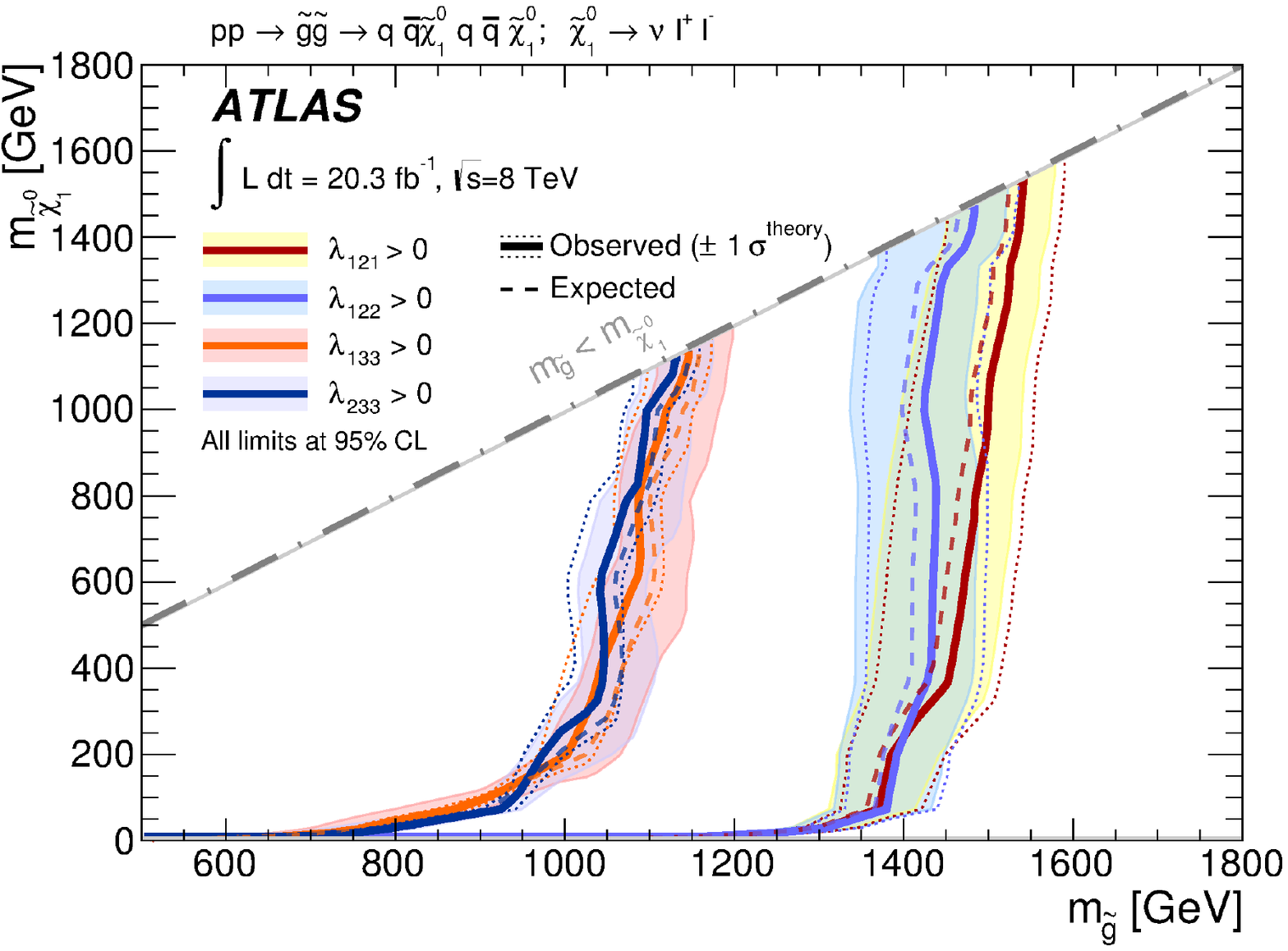}
    \includegraphics[width = 0.29 \textwidth]{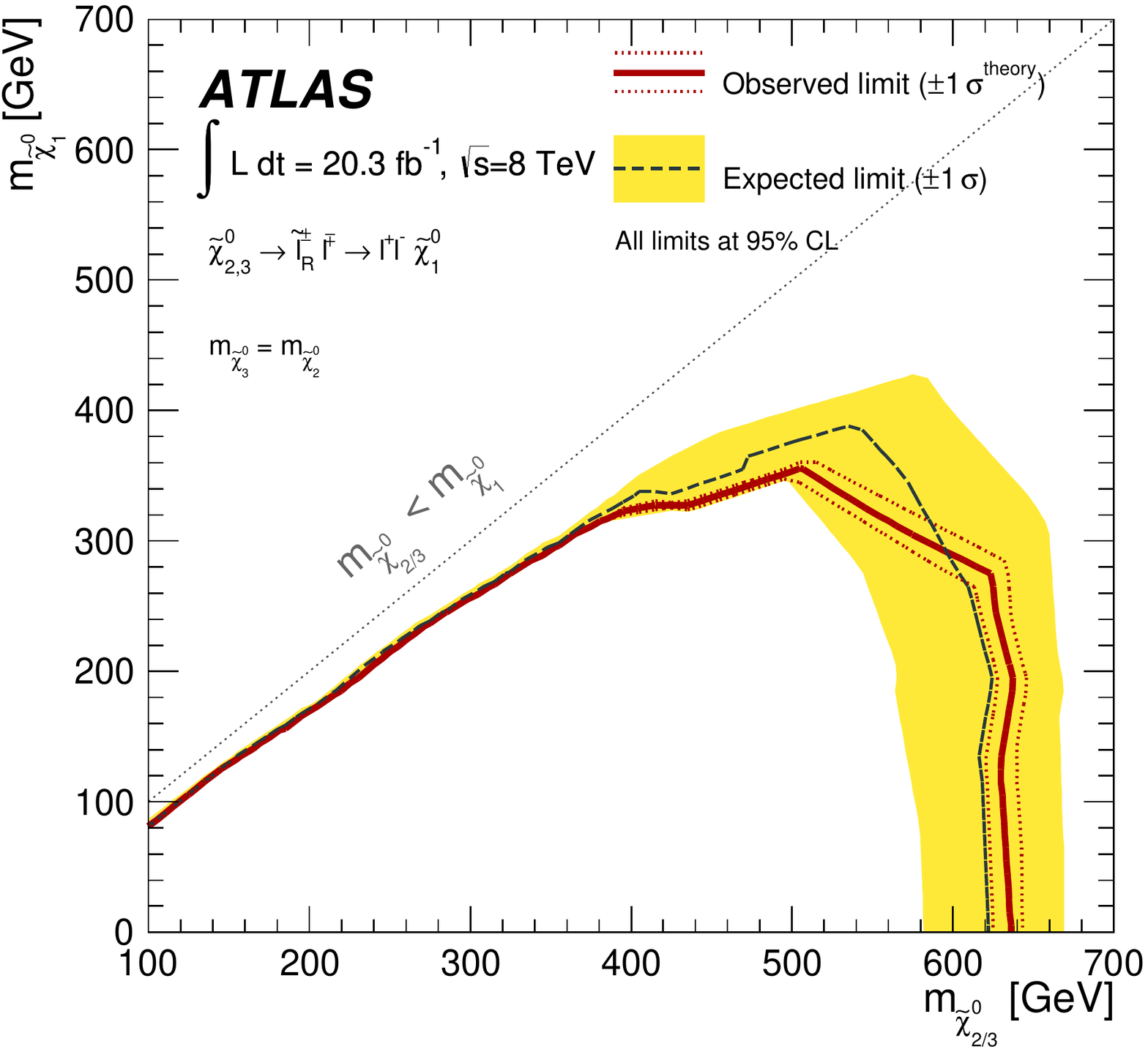}
    \includegraphics[width = 0.29 \textwidth]{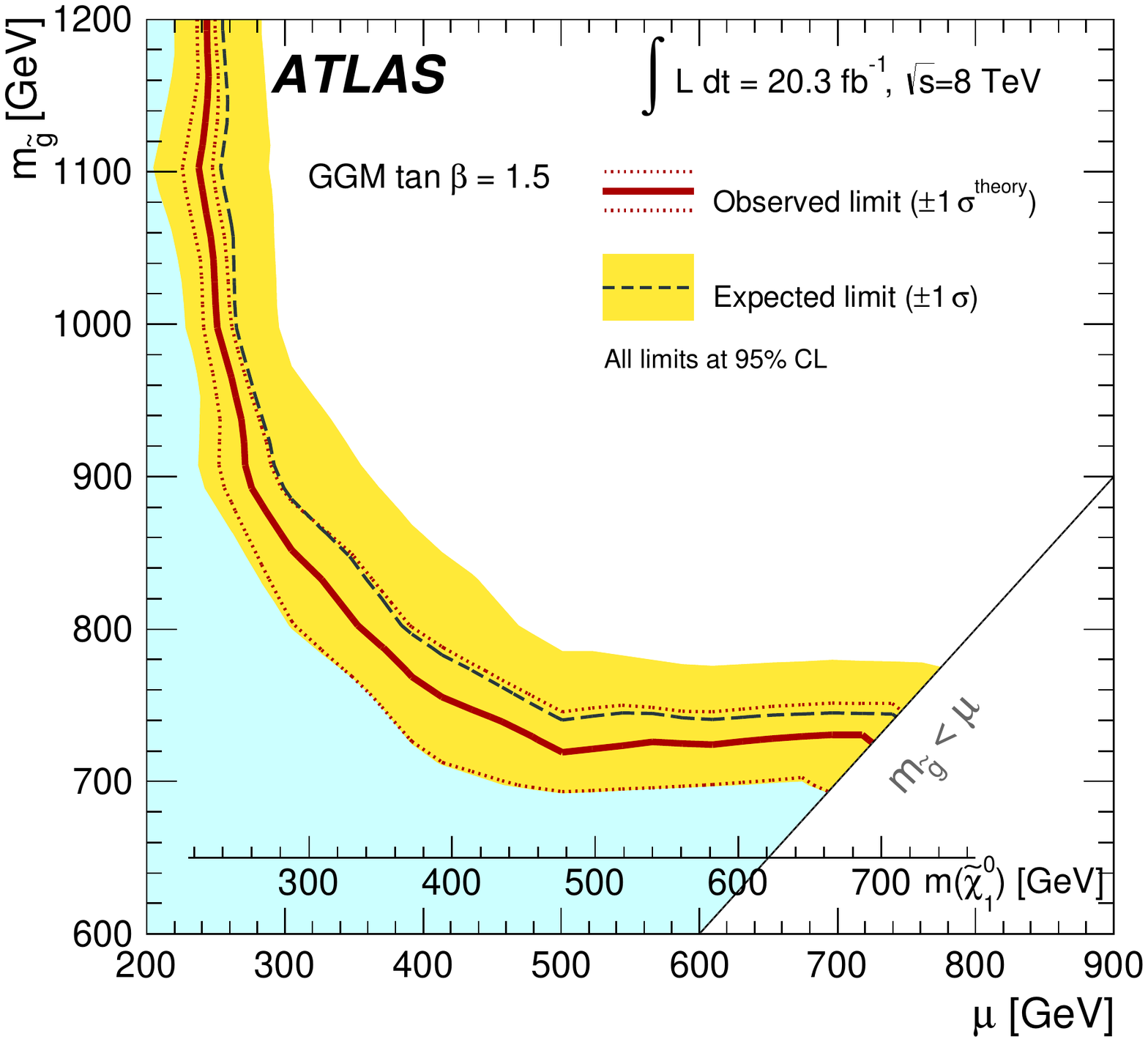}
    \caption{Expected and observed exclusion limits for RPV Gluino pair production (left), RPC Neutralino pair production (center) and the GGM model with $\tan \beta$ = 1.5 (right)\cite{Aad:2014iza}. \label{Limits}}
\end{figure}

\section{Conclusions}

The four-lepton final state is sensitive to a range of different supersymmetric signals. In the search carried out by ATLAS in the full 2012 dataset taken at $\sqrt{s} = 8$ TeV, no excess over the standard model is observed. 
In an R-Parity violating Gluino pair production model, Gluino masses of less than $m(\tilde{g}) < 1350$ GeV are excluded for RPV couplings only resulting in light leptons. If the opposite case of couplings leading to $\tau$ leptons is assumed, the exclusion reduced to $m(\tilde{g}) < 950$ GeV. 
In an RPC model involving electroweak production of heavy higgsino-like $\tilde\chi_{2}^0 \tilde\chi_{3}^0$ pairs to a bino-like $\chi_{1}^0$ via intermediate R-Sleptons, masses of the heavy neutralinos of up to 620 GeV are excluded under the assumption of a massless lightest neutralino. 
In a GGM model with $\tan \beta = 1.5$, values of $\mu$ between 200 GeV and 230 GeV are excluded for any gluino mass, and gluino masses below $m(\tilde{g}) < 700$ GeV are excluded independently of the $\mu$ parameter.



\end{document}